# Language Independent Emotion Quantification using Non-linear Modelling of Speech

Uddalok Sarkar[a,e], Sayan Nag[a], Chirayata Bhattacharyaa[a], Shankha Sanyal[a,f],
Archi Banerjee[a,b,c], Ranjan Sengupta[a] and Dipak Ghosh[a]
[a]Sir C.V. Raman Centre for Physics and Music, Jadavpur University
[b] Department of Humanities and Social Sciences, IIT Kharagpur
[c]Shrutinandan School of Music, Kolkata
[d]Department of Physics, Jadavpur University
[e]Department of Electrical Engineering, Jadavpur University
[f]School of Languages and Linguistics, Jadavpur University

**ABSTRACT**
*At present emotion extraction from speech is a very important issue due to its diverse applications. Hence, it becomes absolutely necessary to obtain models that take into consideration the speaking styles of a person, vocal tract information, timbral qualities and other congenital information regarding his voice. Our speech production system is a nonlinear system like most other real world systems. Hence the need arises for modelling our speech information using nonlinear techniques. In this work we have modelled our articulation system using nonlinear multifractal analysis. The multifractal spectral width and scaling exponents reveals essentially the complexity associated with the speech signals taken. The multifractal spectrums are well distinguishable the in low fluctuation region in case of different emotions. The source characteristics have been quantified with the help of different non-linear models like Multi-Fractal Detrended Fluctuation Analysis, Wavelet Transform Modulus Maxima. The Results obtained from this study gives a very good result in emotion clustering.*

## INTRODUCTION

Emotion is a psychological state related with the sensory system brought on by hormonal changes differently connected with contemplations, sentiments, social reactions, and a level of joy or disappointment. This hormonal changes cause audibly distinguishable features in speech and visibly distinct features in facial expressions. This audio-visual communication helps people to perceive each other's emotion. Audio signal, including both speech and music, in general are inherently complex and few non-linear studies have been conducted to understand the underlying complexity and non-linearity of these signals and their manifestation in the human brain [1-12]. This underlying complex nature of audio signals makes emotion recognition a troublesome task for a machine. The motivation behind emotion recognition framework is to utilize feeling related learning so that human machine correspondence will be improved [13].

Several earlier works dealt with audio-visual data for emotion recognition.[14]. In our study we are concerned with quantification of emotions from speech audio signals. During the past few decades, emotion recognition systems are generally dependent on the very popular speech features like, linear prediction cepstrum coefficient (LPCC), Mel-frequency cepstrum coefficient (MFCC) etc., which are the basis of speech signal processing even today. But these correspond to linear modelling of speech, for e.g., linear predictive coefficients (LPC) assumes vocal tract to be a linear time varying system, in which the present output linearly depends on past outputs and input glottal pulse [15]. But speech production system is not a linear system as is most other physical systems around the world. While using linear modelling of speech, numerous high frequencies are left unattended which may have significant contributions in emotion establishment [16].

In speech production system, while uttering a voiced phoneme (which is contained in most speech segments) air puffs from lungs resonates the vocal fold and obtains quasi-periodic oscillations which is then fed to the time-varying vocal tract to obtain different voiced segments. Both these processes of quasi-periodic oscillation formation and voiced segment formation are non-linear in nature.

Here comes the relevance of non-linear dynamical modelling speech production system, i.e., non-deterministic /chaotic approaches in understanding the speech signals [17-19]. Fractal analysis of audio signals was first performed by Voss and Clarke [20], who analyzed amplitude spectra of audio signals to find out a characteristic frequency f c , which separates white noise (which is a measure of flatness of the power spectrum) at frequencies much lower than f c from very correlated behavior ($1/f^2$) at frequencies much higher than f c . It is well-established experience that naturally evolving geometries and phenomena are rarely characterized by a single scaling ratio; different parts of a system may be scaled differently. That is, the clustering pattern is not uniform over the whole system. Such a system is better characterized as 'multifractal' [21]. Speech production is thus a multifractal phenomenon. There are several methods for measuring fractal dimension, including Detrended fluctuation Analysis, Multi-fractal Detrended fluctuation Analysis, Power Spectral Visibility Graph, Wavelet Transform Modulus Maxima [22], etc. In this work, we have used wavelet transform modulus maxima (WTMM) and Multi-fractal Detrended fluctuation Analysis (MFDFA) for analyzing speech samples and recognition of three different emotions, viz. Happiness, Neutral and Sad.

**EXPERIMENTAL DETAILS**

In our experiment we have tried to establish a robust nonlinear model of emotion estimator which is capable of recognizing the emotions from different languages. So we chose two different languages in this experiment, i.e., English and German. We have chosen The Berlin Database of Emotional Speech [23] from Technical University Berlin for German language and the TESS collection from University of Toronto Psychology Department [24] for English Language.

Berlin Database is recorded in an anechoic chamber of the Technical University Berlin, department of Technical Acoustics under Professor Dr. W. Sendlmeier, Technical University of Berlin, Institute of Speech and Communication, department of communication science. Ten actors (5 men and 5 women) tried to emulate ten sentences in seven different emotions. The emotions taken are neutral (neutral), anger (Ärger), fear (Angst), joy (Freude), sadness (Trauer), disgust (Ekel) and boredom (Langeweile). The total number of existing clips are 800 (number of speakers * number of sentences * number of emotions + some second versions). Recordings were taken with sampling frequency of 48 KHz and thereafter down-sampled to 16 KHz.

The TESS collection from University of Toronto Psychology Department consists of 2800 voice samples of a 26 year old subject and a 64 year old subject. In this database they have considered six different emotions, neutral, anger, fear, happiness, sadness, disgust. A set of 200 target words are spoken in English in the carrier phrase 'Say the word ----' by the two actresses (subjects). The clips are digitized with a sampling frequency of 24414 Hz and are of type mono and in .wav format.

**METHOD OF ANALYSIS**

**Method of Wavelet Transform Modulus Maxima**

In case of measuring Multifractality Wavelet Transform Modulus Maxima or WTMM has been a very well-known method. The time series data obtained from the sound signals are analyzed using MATLAB [24]. Let the obtained discretized signal is $x_k$ of length $N$. Then the wavelet transform can be defined as,

$$W(n,s) = \frac{1}{s}\sum_{k=1}^{N} x_k \, \psi\left(\frac{k-n}{s}\right) \qquad (1)$$

Where $\psi\left(\frac{k}{s}\right)$ the analyzing wavelet is function and $s$ be the corresponding scale parameter. Wavelets are chosen orthogonal to the possible trend, i.e., if the trend is represented by a polynomial, then m-th derivative of a Gaussian is a good choice of mother wavelet so that transform can eliminate trends up to (m-1) order [21]. Now for a certain scaling parameter $s$ the positions $n_i$'s of local maxima of $|W(n,s)|$, is found. Now the q-th power of these modulus maxima are summed over the $n_i$'s for a certain scale $s$

$$Z(q,s) = \sum |W(n_i,s)|^q \qquad (2)$$

The scaling behavior of $Z(q,s)$ is observed to be $Z(q,s) \sim s^{\tau(q)}$. Where the exponents $\tau(q)$ characterize the multifractality of the signal under consideration.

**Method of multifractal analysis of sound signals**

MFDFA is a more generalized version of Detrended fluctuation Analysis or DFA. But a real world time series is not monofractal, in most of the cases long range correlations on small scales can be found which are unattended by any other linear models and DFA process. Here comes the significance of multifractality, hence MFDFA. The entire generalized MF-DFA process is subdivided into five steps. The first three steps are same as DFA. For each step, an equivalent mathematical representation is given which is taken from the prescription of Kantelhardt et al [21].

The complete procedure is divided into the following steps:

*Step 1:* The 'Profile' of the time series is determined. It can be represented as:

$$Y(i) = \sum (x_k - \bar{x}) \quad (3)$$

Where $\bar{x}$ is the mean value of the signal.

*Step 2:* The whole length of the signal is divided into $Ns$ no of segments for different scale parameter $s$. For $s$ as sample size and N the total length of the signal the number of segments are

$$Ns = int\left(\frac{N}{s}\right) \quad (4)$$

*Step 3:* The local RMS variation for any sample size $s$, $F(s,v)$ is determined as follows:

$$F^2(s,v) = \frac{1}{s}\sum_{i=1}^{s}\{Y[(v-1)s+i] - y_v(i)\}^2 \quad (5)$$

*Step 4:* The q-order overall RMS variation for various scale sizes can be obtained by the use of following equation

$$F_q(s) = \left\{\frac{1}{Ns}\sum_{v=1}^{Ns}[F^2(s,v)]^{\frac{q}{2}}\right\}^{\left(\frac{1}{q}\right)} \quad (6)$$

*Step 5:* The scaling behavior of the fluctuation function is obtained by drawing the log-log plot of $F_q(s)$ vs. s for each value of q.

$$F_q(s) \sim s^{h(q)} \quad (7)$$

The h(q) is called the generalized Hurst exponent. The Hurst exponent is measure of self-similarity and correlation properties of time series produced by fractal. The presence or absence of long range correlation can be determined using Hurst exponent. A monofractal time series h(q) is independent of q since the local RMS variance $F^2(s,v)$ will be identical for all segments. The relation between generalized Hurst exponent h(q) of MFDFA and classical scaling exponent τ(q) is defined by the relation

$$\tau(q) = qh(q) - 1 \quad (8)$$

The singularity spectrum f(α) is related to h(q) by a Legendre Transform [13, 14]

$$\alpha = \tau'(q)$$
$$\Rightarrow \alpha = h(q) + qh'(q) \quad (9)$$
$$f(\alpha) = q\alpha - \tau(q)$$
$$\Rightarrow f(\alpha) = q[\alpha - h(q)] + 1 \quad (10)$$

Where α denoting the singularity strength and $f(\alpha)$, the dimension of subset s characterized by α. The spectra can be characterized quantitatively by fitting a quadratic function with the help of least square method [21] in the neighbourhood of maximum $\alpha_0$,

$$f(\alpha) = A(\alpha - \alpha_0)^2 + B(\alpha - \alpha_0) + C \quad (11)$$

Here C is an additive constant C = f(α₀) = 1and B is a measure of asymmetry of the spectrum. The width of the multifractal spectrum essentially denotes the range of exponents. We can obtain the width of the spectrum very easily by extrapolating the fitted quadratic curve to zero, i.e., Width W $W = \alpha_1 - \alpha_2$ with $f(\alpha_1) = f(\alpha_2) = 0$.
The width of the spectrum gives a measure of the multifractality of the time series. For a monofractal time series, the width will be zero as h(q) is independent of q and for higher Multifractal width the time series will be strongly multifractal in nature. The spectral width and spectral orientation has been considered as a parameter to evaluate how the features of speech of a particular emotion varies from another.

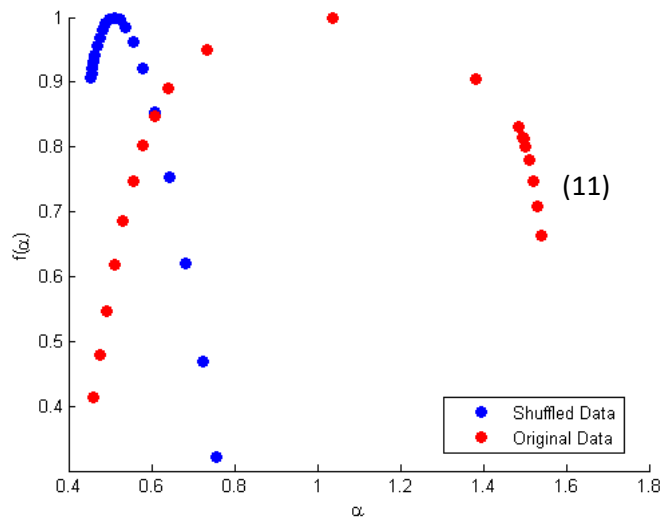

**Figure 1.** MF-DFA spectrum of shuffled data and original data

All the speech samples corresponding to different languages were randomly shuffled before they were put to test using the two methodologies referred to here. The main aim of this work is to develop an automated speech emotion classifier which is independent of any spoken language used.

## RESULTS AND DISCUSSION

The multifractality in speech results in distinct features and properties in case of different emotions. All the speech samples taken were analyzed both with WTMM and MFDFA technique and the multifractal spectra was plotted in **Fig. 2 and 3** for both the methodologies. As a pilot study, we chose to classify two basic emotions i.e. happiness and sadness, and compare them with a neutral emotion using the two methodologies described above - WTMM and MFDFA. As in case of Wavelet transform modulus maxima, distinct categorization is noted for the different emotional speeches. From **Fig.2,** one can notice significant clustering of emotional speeches corresponding to their Holder exponent values. While happy emotional speeches pertain to the low Holder exponent region, sad emotional speeches cluster in the high Holder exponent region. The neutral speeches lie somewhere in between. The multifractal spectral width from MFDFA method reveals essentially the complexity associated with the speech signals taken. From **Fig. 3,** it can be seen that the multifractal width is always greater than 0.5, hence it can be inferred that all the speech signals analyzed are strongly multifractal in nature. The origin of multifractality in most cases is due to the long range fluctuations and broad probability distribution of the source signals. From **Fig. 3** distinct clustering of multifractal spectra is obtained for each emotional attribute mainly in low fluctuation region. For lower Holder's exponents, the singularity spectrum $f(\alpha)$ has different spectral orientations for different emotions, whereas for higher Holder's exponent, i.e., in high fluctuation region significant overlap is noted. Hence, it is difficult to draw any conclusive inference from the high fluctuation region.

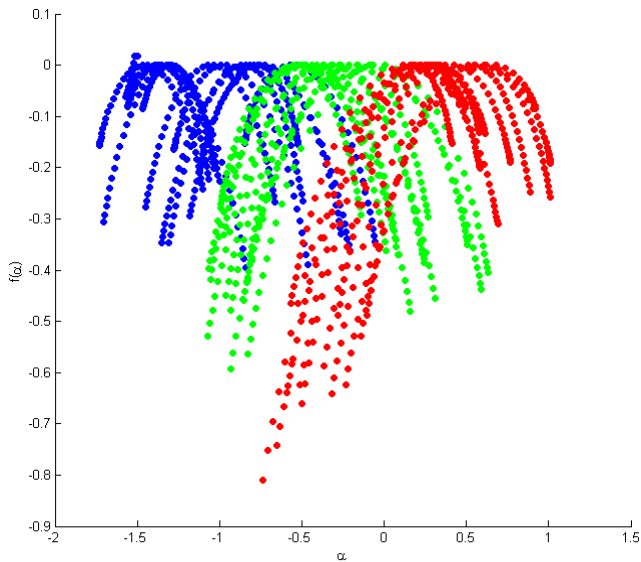

**Figure 2**. WTMM spectra. Blue spectrums corresponds to Happiness, Green Corresponds to Neutral and Red corresponds to Sadness.

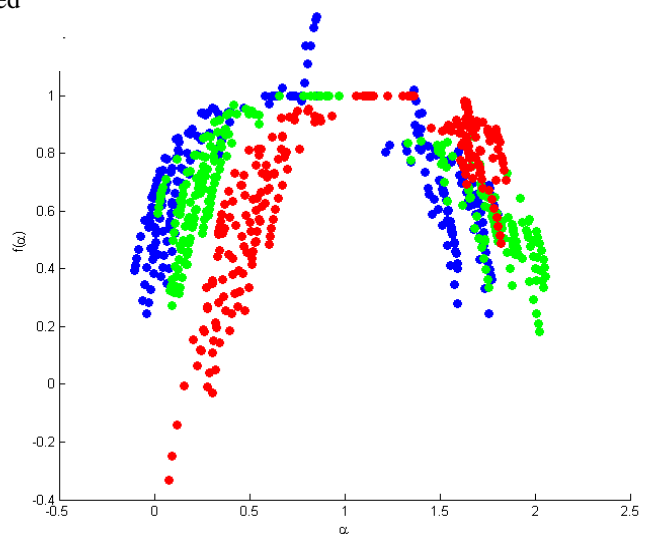

A glance at both the figures reveal that WTMM method provides much better classification accuracy of emotions as compared to MFDFA method. Although in both the methods, the segregation is much more prominent in the low fluctuation region. Thus in any future work related to emotion recognition and classification from speech samples, the low fluctuation region of multifractal spectrum should be the region of interest.

**Figure 3.** MFDFA spectra; Blue spectrums corresponds to Happiness, Green Corresponds to Neutral and Red corresponds to Sadness.

For the purpose of classification, the following statistical features have been used:
1. Features from MFDFA spectrum: In case of MFDFA spectrum we calculate the area under the spectrum for the low fluctuation region. We make use of the following relation to derive the are under the curve -

$$S = \int_{\alpha_{min}}^{\operatorname{argmax}_\alpha f(\alpha)} f(\alpha)d\alpha$$

Here $\operatorname{argmax}_\alpha f(\alpha)$ defines the upper bound of low fluctuation region.

2. Features from WTMM spectrum: In case of WTMM spectrum we extrapolate the fitted quadratic curve to zero. Thus, we find two values of $\alpha$, say $\alpha_1$ and $\alpha_2$ such that $f(\alpha_1) = f(\alpha_2) = 0$ where $\alpha_1$ denotes the zero in low fluctuation region and $\alpha_2$ denotes the zero in high fluctuation region.

Thus, we find a three-dimensional feature space with ($S$, $\alpha_1$, $\alpha_2$). After extracting these features for the whole speech database we fitted it with support vector machine. Using linear kernel we got an average accuracy of 96.3% with a standard deviation of 2.621.

**Fig.4** shows a confusion matrix of a classification run which gives an accuracy of 97.2%. In this run 12 random clips from respective emotions are taken. We can see all the clips from Neutral emotion and sadness have been classified perfectly. One clip from Happiness is classified as Neutral. This case can be justified by the slight overlapping region between Neutral and Happiness from the MFDFA and WTMM spectrums.

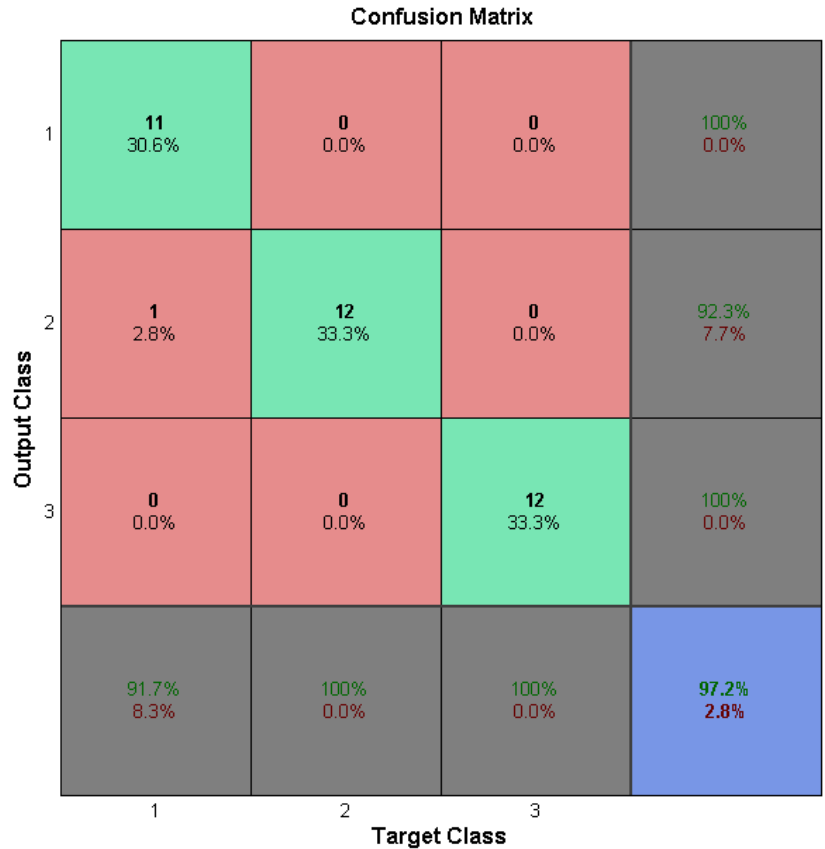

**Fig.4:** Confusion matrix of one classification run with an accuracy of 97.2 % (Here class 1 defines Happiness, class 2 defines Neutral emotion and class 3 defines Sadness)

**CONCLUSION**

In this study, we propose a novel emotion recognition model with which emotions originating from speech signals can be classified cross-linguistically and also compared two methodologies for doing so. The main findings from the study can be listed as under:

1. A language independent robust emotion classifier algorithm can be developed using robust fractal methodologies, which takes into account the inherent non-stationary fluctuations of the speech signals.
2. In both the methodologies used, it is seen that distinct clustering of two strongest emotions - happy and sad is noted mainly in the low fluctuation region.
3. WTMM works more efficiently in clustering emotions from speeches as compared to MFDFA.
4. Using different statistical features from the spectrums, an average classification accuracy of 96.3% is obtained for the methods employed.

This is a pilot study in developing language independent speech emotion classifier which is being further expanded with other languages as well as taking into consideration other major emotions to decipher more interesting results.

**REFERENCES**
1. Bhattacharyya, Chirayata, et al. "From Speech to Recital--A case of phase transition? A non-linear study." arXiv preprint arXiv:2004.08248 (2020).

2. Sengupta, Sourya, et al. "Emotion specification from musical stimuli: An EEG study with AFA and DFA." 2017 4th International Conference on Signal Processing and Integrated Networks (SPIN). IEEE, 2017.
3. Nag, Sayan, et al. "Can musical emotion be quantified with neural jitter or shimmer? A novel EEG based study with Hindustani classical music." 2017 4th International Conference on Signal Processing and Integrated Networks (SPIN). IEEE, 2017.
4. Sanyal, Shankha, et al. "Music of brain and music on brain: a novel EEG sonification approach." Cognitive neurodynamics 13.1 (2019): 13-31.
5. Sarkar, Uddalok, et al. "Speaker Recognition in Bengali Language from Nonlinear Features." arXiv preprint arXiv:2004.07820 (2020).
6. Sarkar, Uddalok, et al. "A Simultaneous EEG and EMG Study to Quantify Emotions from Hindustani Classical Music." Recent Developments in Acoustics. Springer, Singapore, 2020. 285-299.
7. Sanyal, Shankha, et al. "Tagore and neuroscience: A non-linear multifractal study to encapsulate the evolution of Tagore songs over a century." Entertainment Computing (2020): 100367.
8. Banerjee, Archi, et al. "A novel study on perception–cognition scenario in music using deterministic and non-deterministic approach." Physica A: Statistical Mechanics and its Applications 567 (2021): 125682.
9. Banerjee, Archi, et al. "Neural (EEG) Response during Creation and Appreciation: A Novel Study with Hindustani Raga Music." arXiv preprint arXiv:1704.05687 (2017).
10. He, Juan, et al. "Non-Linear Analysis: Music and Human Emotions." 2015 3rd International Conference on Education, Management, Arts, Economics and Social Science. Atlantis Press, 2015.
11. Sanyal, Shankha, et al. "Gestalt Phenomenon in Music? A Neurocognitive Physics Study with EEG." arXiv preprint arXiv:1703.06491 (2017).
12. Sanyal, Shankha, et al. "A Non Linear Approach towards Automated Emotion Analysis in Hindustani Music." arXiv preprint arXiv:1612.00172 (2016).
13. Chiriacescu, "Automatic Emotion Analysis Based On Speech", M.Sc. THESIS Delft University of Technology, 2009
14. Zeng, Z., Pantic, M., Roisman, G. I., & Huang, T. S. (2008). A survey of affect recognition methods: Audio, visual, and spontaneous expressions. *IEEE transactions on pattern analysis and machine intelligence*, *31*(1), 39-58.
15. Pan, Yixiong, Peipei Shen, and Liping Shen. "Speech emotion recognition using support vector machine." *International Journal of Smart Home* 6.2 (2012): 101-108.
16. Bhaduri, S., and D. Ghosh. "Non-invasive detection of alzheimer's disease—multifractality of emotional speech." *J. Neurol. Neurosci* (2016).
17. Behrman, A. (1999). Global and local dimensions of vocal dynamics. *JOURNAL-ACOUSTICAL SOCIETY OF AMERICA*, *105*, 432-443.
18. Kumar, A., & Mullick, S. K. (1996). Nonlinear dynamical analysis of speech. *The Journal of the Acoustical Society of America*, *100*(1), 615-629.
19. Sengupta, R., Dey, N., Nag, D., & Datta, A. K. (2001). Comparative study of fractal behavior in quasi-random and quasi-periodic speech wave map. *Fractals*, *9*(04), 403-414.
20. Voss, R. F., and J. Clarke. "1/f noise in speech and music." *Nature* 258 (1975): 317-318.
21. Kantelhardt, J. W., Zschiegner, S. A., Koscielny-Bunde, E., Havlin, S., Bunde, A., & Stanley, H. E. (2002). Multifractal detrended fluctuation analysis of nonstationary time series. *Physica A: Statistical Mechanics and its Applications*, *316*(1-4), 87-114.
22. Muzy, Jean-François, Emmanuel Bacry, and Alain Arneodo. "Multifractal formalism for fractal signals: The structure-function approach versus the wavelet-transform modulus-maxima method." *Physical review E* 47.2 (1993): 875.
23. Burkhardt, Felix, et al. "A database of German emotional speech." *Ninth European Conference on Speech Communication and Technology*. 2005.
24. Ihlen, E. A. F. E. (2012). Introduction to multifractal detrended fluctuation analysis in Matlab. *Frontiers in physiology*, *3*, 141.
25. Feder, J. "Fractals. Plenum Press, New York." *Fractals. Plenum Press, New York.* (1988).
26. Peitgen, Heinz-Otto, H. Jurgens, and D. Saupe. "Chaos and fractal." *New Frontiers of Science, Springer Verlag* (1992).